\documentclass{article} % For LaTeX2e
\usepackage{nips15submit_e,times}
\usepackage{hyperref}
\usepackage{url}

\usepackage{graphicx}
\usepackage{caption,subcaption}
\usepackage{multirow}
\usepackage{tabulary}
\newcolumntype{K}[1]{>{\centering\arraybackslash}p{#1}}
\usepackage{listings}
\usepackage[export]{adjustbox}[2011/08/13]
\usepackage{fancyhdr}
\fancypagestyle{firststyle}{
    
    \fancyfoot[L]{A poster with the same title was presented at the 2017 International Symposium on Field Programmable Gate Arrays}
}

\def\ctext/#1{{\ttfamily{#1}}}
\def\mypar/#1{{\bf{#1.}}}

\title{Using Vivado-HLS for Structural Design: \protect\\
a NoC Case Study}

\author{
Zhipeng Zhao \\
ECE Department\\
Carnegie Mellon University\\
Pittsburgh, PA 15213 \\
\texttt{zzhao1@andrew.cmu.edu} \\
\And
James C. Hoe \\
ECE Department\\
Carnegie Mellon University\\
Pittsburgh, PA 15213 \\
\texttt{jhoe@ece.cmu.edu} \\
}

\nipsfinalcopy % Uncomment for camera-ready version

\begin{document}

\begin{center}
\end{center}

\maketitle
\thispagestyle{firststyle}
\begin{abstract}

There have been ample successful examples of applying Xilinx Vivado's
``function-to-module'' high-level synthesis (HLS) where the subject is
algorithmic in nature.  In this work, we carried out a design study to assess
the effectiveness of applying Vivado-HLS in structural design.  We
employed Vivado-HLS to synthesize C functions corresponding to
standalone network-on-chip (NoC) routers as well as complete
multi-endpoint NoCs.  Interestingly, we find that describing a complete
NoC comprising router submodules faces fundamental difficulties
not present in describing the routers as standalone modules. 
Ultimately, we
succeeded in using Vivado-HLS to produce router and NoC modules that are
exact cycle- and bit-accurate replacements of our reference RTL-based
router and NoC modules.  Furthermore, the routers and NoCs resulting from
HLS and RTL are comparable in resource utilization and critical path
delay.  Our experience subjectively suggests that HLS is able to
simplify the design effort even though much of the structural details
had to be provided in the HLS description through a combination of
coding discipline and explicit pragmas. The C++ source code can be found at \url{http://www.ece.cmu.edu/calcm/connect_hls}.

\end{abstract}

\section{Introduction}

Xilinx Vivado offers a style of high-level synthesis (HLS) that maps a
restricted C function to a hardware module, eschewing the question of
supporting the complete C language or mapping a complete C program.
Commercial interests aside, this technology has garnered significant
interest among researchers across domains as evident in the number of recent papers
where Vivado-HLS is used for application development.

In a typical usage, the designer expresses the desired algorithm using
C constructs in the function body.  Such a C function could have many
legal interpretations as a hardware module, differing widely in the
timing and structure of the datapath and interface.  In mapping C code
to a datapath, Vivado-HLS can automatically make default implementation
choices where the C specification is silent.  Alternatively, the
designer has the option to steer many of the mapping decisions using
Vivado-HLS provided pragmas~\cite{Vivado_user_manual}.

In our own work, we have used Vivado-HLS to develop compute kernels
of signal and vision processing algorithms.  We have found the combination of
HLS design, debug, and analysis environments to indeed improve our
productivity without compromising quality relative to
what we could have done using register-transfer-level (RTL) Verilog.
In our uses, although we are working through C and HLS, we
have always approached the design with a skeletal model of the
desired datapath in mind. And, thus far, we have been successful in
coaxing Vivado-HLS into producing the intended datapath---and thus the
expected quality of result---through a combination of pragmas and
rewriting/restructuring of the code.  {\em {This led us to ask the
    question, can we use C and Vivado-style HLS for pure structural
    design?  Moreover, would there be any benefits in using C and
    Vivado-style HLS for structural design?}} \footnote{To be precise,
  we are interested in C/C++ specifications of behaviors that are
  readable as C/C++ programs. Synthesizable SystemC, based on
  processes with sensitivity list, is used as an RTL language in the
  same way as synthesizable Verilog and VHDL.} (Ans: yes and sometimes.)

By structural design, we mean a design that carries strong explicit notions of
synchronous sequential state elements and combinational next-state
logic where the designer {\em{wants}} to control with precision the
design's structure and timing, cycle-by-cycle and bit-by-bit.  The
difference between structural vs. non-structural descriptions is
orthogonal from the style of a language or the language's level of
abstraction.  Synthesizable RTL subsets of Verilog, VHDL, and SystemC
are naturally aligned for structural design descriptions.
Chisel~\cite{chisel_user_manual} and
Bluespec~\cite{Bluespec_user_manual}, which borrow heavily from
modern functional programming languages, are very high-level but still
structural in nature (requiring explicit declarations of synchronous state
variables and their state transition logic).  In this paper, we aim to use C and Vivado-HLS to achieve cycle- and bit-level control of the
desired datapath structure.

To understand the effectiveness of C and Vivado-HLS for structural
design, we chose network-on-chip (NoC) routers as the subject of a
design study.  Except for perhaps the allocation logic, a NoC router
is manifestly ``structural''.  For this study, we set out to replicate
the RTL router designs available through
CONNECT~\cite{IEEE_computer_article}.  To push the limit, we further
attempted to describe complete NoCs with routers encapsulated as
submodules.

We were ultimately successful in using C/C++ and Vivado-HLS to produce router modules that are exact cycle- and bit-accurate replacements for
the full range of diversely parameterized CONNECT family of routers.
Moreover, the routers synthesized by Vivado-HLS are comparable with the quality of corresponding CONNECT routers.

At the scope of a router, C and Vivado-HLS proved to be competent in
capturing the structural intents and producing the desired outcome.
More importantly, we found that the HLS design flow allowed us to
address functional correctness separately from structural
decisions.  Relative to traditional RTL design flow, we were able to
more easily and quickly achieve functional correctness due to C's
sequential semantics and its development/debug environment.
Afterwards, we could also more rapidly explore different structural
design alternatives without worrying about breaking functional
correctness.  This decoupling of the functionality and structural
aspects of a design resulted in a clearly perceivable productivity
gain, even for this clearly structural design effort where the C and
corresponding RTL routers descriptions must carry the same information
with respect to the design's structure and timing.

On the other hand, when expanding the design scope to a NoC, we saw
great interference from the semantics of C in creating a ``netlist''
of router submodule instances.  We developed a solution
involving describing the routers as C++ objects under an imposed
coding discipline.  All in all, our experience suggests that C/C++ and
Vivado-HLS can be used to replicate arbitrary RTL designs of
synchronous registers and combinational logic, but the benefits in
doing so depend on the hierarchical nature of the design subject.

\mypar/{Outline} Following this introduction,
Section~\ref{sec:background} offers a brief background review of HLS
and the structure of CONNECT routers.  Section~\ref{sec:hls} explains
how one could describe a router-like structure in C for
HLS. Section~\ref{sec:noc} next discusses the challenges when
expanding the design scope from individual routers to complete NoCs.
Section~\ref{sec:tuning} discusses the limiting cases in using C and
Vivado-HLS for structural design.  Section~\ref{sec:results} presents
the place-and-route evaluation results comparing HLS routers and NoCs
with their CONNECT counterparts.  Section~\ref{sec:conclusion} offers
our conclusions.

\section{Background}
\label{sec:background}

\subsection{High-Level Synthesis}

The rapidly growing interest in using FPGAs for compute acceleration
has also boosted the interest in HLS as the path to simpler
application development and a larger developer pool.  Historically,
 HLS has not been exclusive to the FPGA domain and
predates FPGAs~\cite{mcfarland1990}.  Among the many abstractions and
languages considered for high-level or behavioral specification, there
has always been a strong emphasis on C for its practical simplicity
and immense popularity.  Over time, the range of work in C-to-hardware
synthesis has spanned from those that only borrowed C's syntax for an RTL
language (e.g., Perle1DC~\cite{active_memory}) to those that compiled complete
C language programs for execution as hardware
(e.g., CASH~\cite{spatial_computation}).  At the same time, arguments have
been raised against the suitability of C---an untimed, sequential
language---for synthesizable hardware description~\cite{Edwards2006}.

To reduce the designer's workload, a behavioral specification is
intentionally under-constrained to defer structural details and
decisions to automation.  As such, a given behavioral specification
could have multiple correct implementations, varying in their datapath
structure and/or timing, with corresponding consequences in their
design quality of concern (performance, area, power, etc.).  For
some behaviors, it is possible that none of the implementations are
``good'' relative to alternatives such as executing in software.  The
great challenge for HLS is then in correctly filling in the missing
details and decisions to arrive at a good implementation when one
exists.  Windh~et~al. offers a survey of the state of the
art in HLS~\cite{Windh2015ProceedingsIEEE}.

The HLS problem can be simplified by restricting the applicable inputs
and/or the range of output structural options.  For example, ROCCC
specializes on mapping parallelizable loop nests to streaming
pipelines~\cite{roccc}.  Such specialization allows
very high quality results for the intended/anticipated usage scenarios.

To support the complete C language and full programs, one could
restrict the language constructs and the program regions
synthesized to hardware. LegUp~\cite{legup} and SDSoC~\cite{sdsoc}
support a hybrid execution where the main program thread is compiled
for execution on an embedded processor core.  Only specially
designated functions---presumably those that make sense to be in
hardware---are synthesized to hardware modules.
The compiler automatically inserts the required
hardware and software interfaces; therefore 
calls to the hardware accelerated functions are
transparent at the source code level.
LegUp can also make use of
threading semantics to support multiple concurrent, free-running
hardware modules~\cite{LegUpStreaming}.

Vivado-HLS, descending from AutoPilot~\cite{CongXilinxpaper}, limits
itself to only the problem of mapping restricted C functions to
modules.  Moreover, Vivado-HLS selectively gives emphasis to
language constructs and code structures that are most important to
hardware-friendly algorithms. For example, Vivado-HLS does not support
recursion and only optimizes loops with fixed loop bounds.
Lastly, Vivado-HLS relies on an extensive suite of pragmas for
designers to explicitly add structural and timing details when the
compiler's default outcome is less than desired.  As mentioned in the
introduction, in the domain of signal and vision processing
algorithms, our own work has used C and Vivado-HLS to good effect in arriving
at the intended high-quality datapath by exercising strong structural
control through code styles and explicit pragmas.  This experience
motivated our current curiosity in how effective is C and Vivado-HLS
for pure structural design.  Xilinx Application Note XAPP1209 ~\cite{xapp1209} and XAPP1167~\cite{xapp1167} illustrate
how to use Vivado-HLS to develop a streaming structure 
comprising fully pipelined modules that interact only through stylized streaming interfaces.  In this paper,
we show how one could use Vivado-HLS to develop
arbitrary structural designs of synchronous registers and combinational logic.  In prior work, Kapre and Gray have used Vivado-HLS to generate standalone router module for Hoplite, a lightweight FPGA overlay NoC~\cite{hoplite}. Lahti, et al. showed how to develop a cycle-accurate structural module of a I2C bus controller using Catapult C~\cite{CatapultC}.

\subsection{Structure of CONNECT Routers}
\label{sec:connect}

We elected to use NoC routers for this structural design study.  In
particular, we set out to replicate the packet-switched routers
available from the CONNECT NoC design generator~\cite{IEEE_computer_article}.  There is no special significance in choosing routers as
the design subject, except that they are structural.  The key ideas developed in this paper are generalizable beyond routers.

%Contrastingly, focusing the study on routers---which are not representative
%of all things structural---requires us to think harder about the
%generalizability of the findings.

%%Using CONNECT does increase diversity by providing RTL references for
%%a diverse family of router designs for this study.

The CONNECT NoC design generator is parameterized to generate
packet-switched routers from a comprehensive design space.  Basic
router design parameters include the in- and out-degree, flit data width, and flit buffer size. More advanced parameters select major
design options in flow control, virtual channel (VC), and allocation.
At the next level, CONNECT is parameterized to construct a number of
stylized or customized network topologies from routers.  CONNECT is
available through a GUI web-based
portal that produces
synthesizable RTL Verilog.\footnote{http://www.ece.cmu.edu/calcm/connect. Internally, the RTL design generator is coded in BSV and uses
  a Bluespec compiler~\cite{Bluespec_user_manual} to produce Verilog.}
Evaluations have shown CONNECT-generated routers
and NoCs are competitive with high-quality hand-coded RTL
designs~\cite{IEEE_computer_article}.  Although not necessary for this
paper, readers desiring a more thorough background in router and NoC
design can refer to~\cite{Dallybook}.

\begin{figure}
\centering
\includegraphics[width=0.43\textwidth]{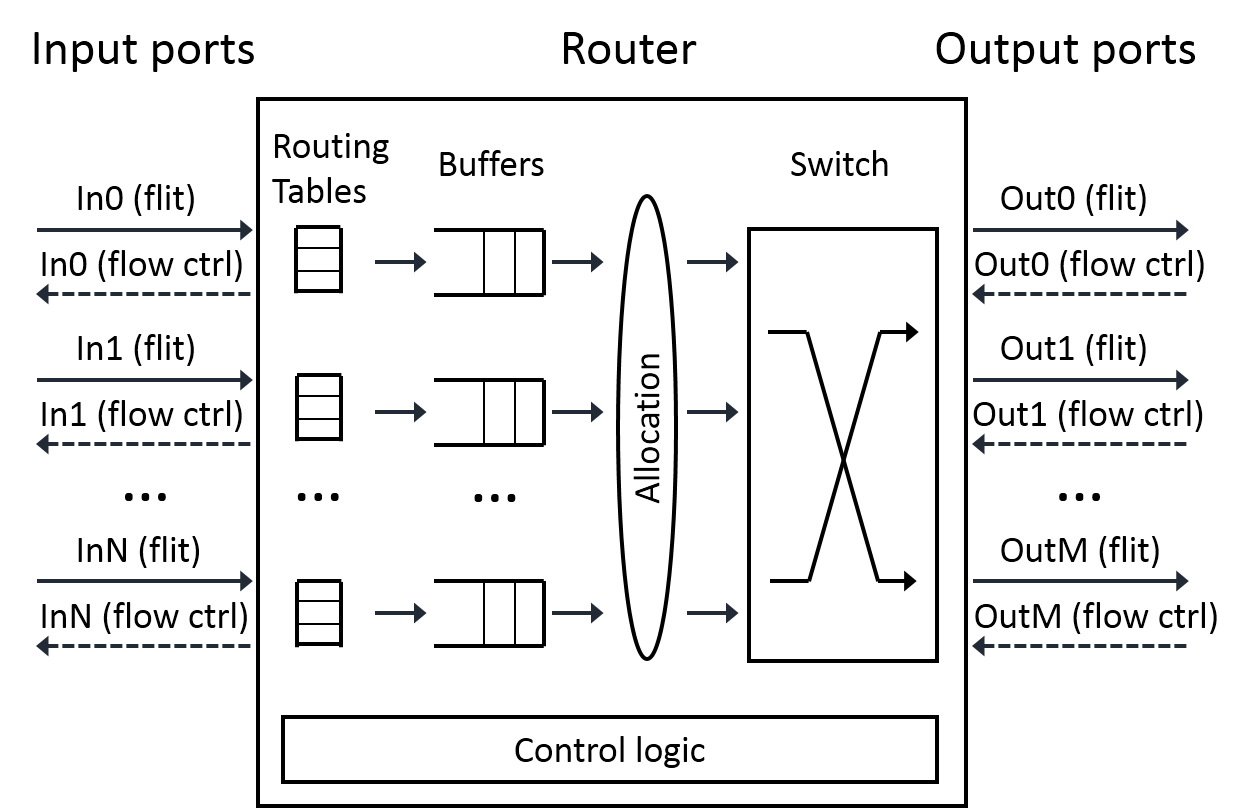}
%\vspace{0pt}
\caption{CONNECT router structural template.}
\label{fig::router}
\end{figure}

The diverse family of CONNECT routers share a common underlying
structure depicted in Figure~\ref{fig::router}.  Generically, a router
is a module with some number of input ports and output ports.  A flit,
comprising data payload bits and route information bits\footnote{It is
  a unique feature of CONNECT routers that each flit of a packet carry
  its own route information.  This is an optimization to
  make use of the excess wiring resources expected when mapping
  regularly tiled structures onto an FPGA fabric~\cite{connect_fpga}.},
can arrive on an input port from time to time.  The job of a router is
to emit the flit eventually on an output port according to the flit's
route information bits.  

Upon entering the router, the flit's
eventual output port is determined combinationally (by a look-up table
for example) using its route information bits. Because the desired
output port is not always immediately available, it is necessary to
first buffer the incoming flit.  As an
optimization (discussed later in Section~\ref{sec:tuning}), the data
payload and the route information portions of the flits can be
physically held in different structures that logically operate in
synchronization as one FIFO. On each cycle, the allocation logic
considers the flits at the front of all FIFO flit buffers to dispatch
a maximum number of flits to their non-conflicting output ports
through a crossbar switch; flits not selected are deferred to the next
cycle.  This datapath from input ports to output ports is pipelined to
increase throughput.  Finally, because an input port's flit buffer can
fill up under output port congestion, an input port needs to
communicate with its upstream router using a flow-control protocol to
ensure the upstream router never overflows the capacity of the flit
buffer.

\begin{figure}
\centering
\includegraphics[width=0.32\textwidth]{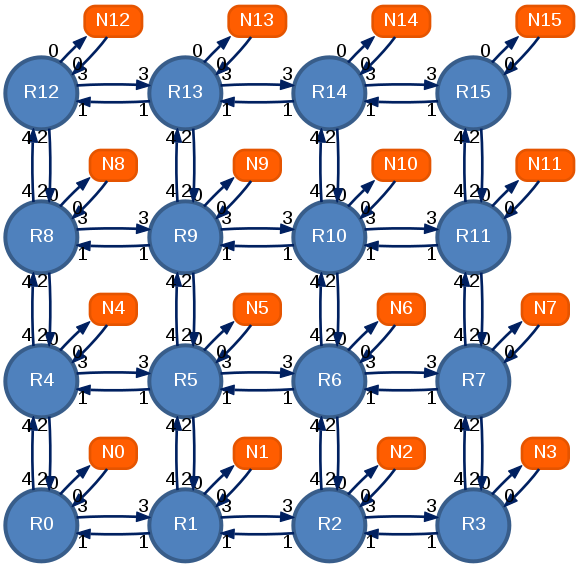}
\vspace{8pt}
\caption{A 4-by-4 2D-Mesh using 5-by-5 routers in the interior, 4-by-4 routers on the edges,  and  3-by-3 routers on the corners.}
\label{fig::2dmesh}
\end{figure}

The basic CONNECT options affect the structure straightforwardly in
the number of ports, the width of the datapath, and the depth of the
buffers.  The advanced CONNECT options primarily affect the combinational
decision logic controlling the datapath. The virtual channel options affect
structure---associating multiple FIFO flit buffers to each input
port---as well as the combinational logic.  In Section~\ref{sec:hls}, we
first describe how C and Vivado-HLS can be used very effectively to
design these kinds of structures.  In Section~\ref{sec:noc}, we next
describe the challenges and solutions in using C and Vivado-HLS to
design NoCs (including,
for example, a 2D-Mesh NoC, Figure~\ref{fig::2dmesh}), using router submodules.

%%Generic Router
%%Architecture. Input flit first goes through routing table, where the
%%target out-port is determined based on the destination information
%%carried by the flit. The first flit in buffer generates request for
%%allocation, and the granted flits will traverse through the
%%switch. Flow control signal is a feedback signal to negotiate with
%%upstream router on flit transmission.}

\def\cI1/{\ctext/{I1}}
\def\I2/{\ctext/{I2}}
\def\Odd/{\ctext/{Odd}}
\def\Even/{\ctext/{Even}}
\def\cL1/{\ctext/{L1}}
\def\L2/{\ctext/{L2}}
\def\Lreq/{\ctext/{Lreq}}
\def\cF1/{\ctext/{F1}}
\def\F2/{\ctext/{F2}}
\def\Oo/{\ctext/{O}}
\def\Ll/{\ctext/{L}}

\section{Describing Structure in C}
\label{sec:hls}

Although we try to be self-contained, this
section is not a good primer for Vivado-HLS.  Our usage is
not orthodox.  We require only a very basic subset of Vivado-HLS
features.

Instead of a router, we use the running example of a simple switch
module that steers the integer inputs on its two input ports \cI1/
and \I2/ toward the appropriate output ports \Odd/ and \Even/.  It
should be possible to completely understand how to describe the
structure of this switch module, and then extrapolate how to
describe the CONNECT router structure discussed in
Section~\ref{sec:connect} \footnote{The source code for the examples in this section and the complete router can
be found at http://www.ece.cmu.edu/calcm/connect\_hls.}. Below, \ctext/{switch\_try()} is a starting attempt at describing
the switch module.\footnote{For clarity, we omit
in the examples the \ctext/{ap\_ctrl\_none} pragma on the functions and
the \ctext/{ap\_none} pragma on the output variables. They serve only to
remove the default control signals that become extraneous in our use.}

\begin{lstlisting}
void switch_try(int I1, int I2, int *Odd, int *Even) {
  if (I1%2)  *Odd=I1;
  else if (I2%2)  *Odd=I2;
  
  if (!(I1%2))  *Even=I1;
  else if (!(I2%2))  *Even=I2;
}
\end{lstlisting}

\mypar/{Mapping Function to Module}
Vivado-HLS maps the top-level function \ctext/{switch\_try()} to a corresponding logic
module. The pass-by-value arguments \cI1/ and
\I2/ correspond to input ports.  The pass-by-reference arguments
\Odd/ and \Even/, only dereferenced for
writing, correspond to output ports.  The body of the function
specifies the calculation of the output values from the input values in each
function invocation.  This simple example---with no cycles or
side-effects in the dataflow between input and output---would by
default map to a combinational module provided the chosen target
clock period is long enough.\footnote{If the chosen target clock period is
too short for the critical path, Vivado-HLS will break the logic into
multiple cycles to meet timing.  The \ctext/{\#pragma HLS LATENCY
max=0} can be added to force Vivado-HLS to ignore the target clock
period restriction.}

%%\footnote{The C language allows
%%many semantically ``correct'' implementations with different execution
%%timing.  Vivado-HLS supports a rich set of pragams to give designer
%%precise control over the timing of the interface (e.g., asynchronous
%%handshake governing the availability of input and outputs) and the
%%execution (e.g., pipelining to overlap the execution of multiple
%%invocations).}

\mypar/{Flow Control}\footnote{Vivado-HLS can automatically introduce select
styles of flow control (e.g., streams). Here, we are interested in the
ability to build arbitrary flow-control protocols explicitly.} The first attempt \ctext/{switch\_try()} is deficient in that if inputs \cI1/ and
\I2/ are both even or odd, input \I2/ is ignored and one
of the output value (\Even/ or \Odd/) is undefined.  The next
attempt \ctext/{switch\_comb()} adds flow-control
handshakes.

\begin{lstlisting}
void switch_comb(vDat I1, vDat I2, 
                   vDat *Odd, vDat *Even,
                   bool *acpt1, bool *acpt2) {
  *acpt1=*acpt2=(*Odd).v=(*Even).v=false;

  if (I1.v && (I1.d%2)) 
     {(*Odd).v=true; (*Odd).d=I1.d; *acpt1=true;} 
  else if (I2.v && (I2.d%2)) 
     {(*Odd).v=true; (*Odd).d=I2.d; *acpt2=true;}
  ... repeat 6~9 for Even ...
}
\end{lstlisting}

The inputs \ctext/{I1} and \ctext/{I2}, and outputs \ctext/{Even} and \ctext/{Odd} have type \ctext/{struct vDat \{bool
    v; int d;\}} where the \ctext/{.v} field is used to indicate the
    validity of the \ctext/{.d} data field.  The function considers
    the inputs' validity in making steering decisions; the function
    also marks the validity of the outputs.  Lastly, the function adds
    two Boolean outputs \ctext/{acpt1} and \ctext/{acpt2} to indicate
    if the corresponding inputs have been accepted.

\mypar/{Sequential State and Pipelining}
The module synthesized from \ctext/{switch\_comb()} remains
combinational by default.  Vivado-HLS could be directed to automatically
synthesize a streaming pipelined module that would overlap multiple
invocations of the function.  In other words, the module would proceed
ahead to accept a new invocation's inputs each cycle with the
corresponding outputs emerging only some pipeline delay later.  This
streaming execution however is not what we need in the current
context.  We want to pipeline the path from \cI1/ and \I2/ to
\Odd/ and \Even/, but we want \ctext/{acpt1}
and \ctext/{acpt2} to remain combinational to qualify the current
inputs \cI1/ and \I2/.  An example of the desired 2-stage pipeline
structure is described explicitly as \ctext/{switch\_2stage()} below.  

\begin{lstlisting}
void switch_2stage(...same as switch_comb...) {
  static vDat L1, L2;
  static Path Lreq;

/* ---- stage 2 ---- */
  Path grnt=allocate(Lreq);
  if (grnt.L1xOdd) *Odd=L1;
  else if (grnt.L2xOdd) *Odd=L2;
  else (*Odd).v=false;
  ... repeat 6~9 for Even ... 

/* ---- stage 1 ---- */
  if (grnt.L1xOdd||grnt.L1xEvn) L1.v=false;
  if (I1.v && (!L1.v)) { *acpt1=true; L1=I1; } 
  else *acpt1=false;
  ... repeat 13~15 for I2 and L2 ...
  
  Lreq=decode(L1,L2);
}
\end{lstlisting}

The synthesized module will execute one invocation of this function to
completion once per cycle. In our use of C and Vivado-HLS, a function
invocation captures the events of a clock cycle, starting with the
combinational propagations based on current state and input values,
ending with the synchronous next-state update.

We can introduce sequential states using static variables that retain
their values across invocations.  The static variables \cL1/, \L2/,
and \Lreq/ are used as pipeline latches (lines 2 and 3).  In any given invocation
of \ctext/{switch\_2stage()}, \cL1/ and \L2/ holds latched
values of \cI1/ and \I2/ from previous invocations.
Another pipeline latch \Lreq/ holds pre-decode connection requests in
\ctext/{struct Path \{bool L1xOdd; bool L1xEven; bool L2xOdd; bool
L2xEven;\}} which is a bitmap of the four possible connections needed by
\cL1/ and \L2/.

The code corresponding to Stage 2 (lines 6$\sim$10) sets the output \Odd/ and
\Even/ based on the pipeline latches \cL1/, \L2/, and \Lreq/.  
For brevity, we assume a combinational function \ctext/{Path
allocate(Path)} exists to compute a bitmap of which requested
connections are granted (line 6).

The code corresponding to Stage 1 (lines 13$\sim$18)---starting only after Stage 2 code
is finished using the old values of \cL1/, \L2/ and \Lreq/---sets the new values of the pipeline latches based on
their old values and current input values.  Please note that the Stage
1 code is written in ``procedural style'' where
\cL1/, \L2/ and \Lreq/ can be read and written multiple times in describing their
``next-state'' values combinationally; only the final values at the end
of the function are latched.  Again for brevity, we assume a
combinational function \ctext/{Path decode(vDat,vDat)} exists to compute a
bitmap of the requested connections (line 18).

\mypar/{Hierarchy and Modularity} This \ctext/{switch\_buffered()} example uses an assumed synthesizable finite FIFO class and \ctext/{switch\_2stage()} in a hierarchical
 structure.  Once again, the synthesized module for \ctext/{switch\_buffered()} will execute one
 invocation to completion once per
 cycle. The declared static FIFO objects \cF1/ and \F2/ will buffer
 the inputs before they are switched onto the outputs.  In lines 6$\sim$10,
 the fronts of the FIFOs are presented
 to \ctext/{switch\_2stage()}. The ``call''
 to \ctext/{switch\_2stage()} sets top-level outputs \Odd/ and \Even/
 directly. In this context, we could transparently replace \ctext/{switch\_2stage()}
 by
\ctext/{switch\_comb()} 
even though they have different timing, because they obey the same flow-control protocol. 
 Inputs \cI1/ and \I2/ are accepted
 into \cF1/ and \F2/, respectively, as long as the FIFOs are not full (lines 12$\sim$14). The front of a FIFO is popped if its value is
 accepted (line 15).  

%At this point, the module corresponding to \ctext/{switch\_buffered}
%has all of the skeletal elements needed a in CONNECT router.

\begin{lstlisting}
void switch_buffered(...same as switch_comb...) { 
  static FIFO<int> F1, F2; 
  bool okX1, okX2;  // combinational "wire" temporaries 
  vDat frontX1, frontX2; // combinational "wire" temporaries
  
  frontX1.v=!F1.empty();  frontX1.d=F1.front();  
  frontX2.v=!F2.empty();  frontX2.d=F2.front();  
  
  switch_2stage(frontX1, frontX2, 
                  Odd, Even, &okX1, &okX2);
  
  if (!F1.full() && I1.v)  
    {F1.push(I1.d); *acpt1=true;}
  else *acpt1=false;
  if (okX1) F1.pop();

  ... repeat 12~15 for I2 and F2 ...
}
\end{lstlisting}

\mypar/{Parameterized Design}
This final example shows how C language facilities can be
used for more maintainable and scalable design capture. Below,
\ctext/{switch\_buffered\_N()} is derived from \ctext/{switch\_buffered()}
for a parameterized number of inputs.  We assume a parameterized
\ctext/{switch\_2stage\_N()} exists.

%\newpage

\begin{lstlisting}
void switch_buffered_N(vDat I[N],
		                   vDat *Odd, vDat *Even,
		                   bool acpt[N]) {
  #pragma  HLS ARRAY_PARTITION variable=I complete dim=1
  #pragma  HLS ARRAY_PARTITION variable=acpt complete dim=1
  
  static FIFO<int> F[N];
  bool okX[N];
  vDat frontX[N];

  for(int i=0;i<N;i++){ 
  #pragma HLS UNROLL
    frontX[i].v=!F[i].empty(); frontX[i].d=F[i].front();
  }
  
  switch_2stage_N(frontX, Odd, Even, okX);
  
  for(int i=0;i<N;i++) {
  #pragma HLS UNROLL
    if (!F[i].full() && I[i].v)  
      {F[i].push(I[i].d); acpt[i]=true;} 
    else acpt[i]=false;
    if (okX[i]) F[i].pop();
  }
}
\end{lstlisting}

In \ctext/{switch\_buffered\_N()}, the inputs have been declared as an
array \ctext/{I[N]}. Similarly, the per-input flow-control
signal \ctext/{acpt} needs to be an array.  We include the
necessary \ctext/{HLS ARRAY\_PARTITION} pragma (lines 4 and 5) to
instruct Vivado-HLS to synthesize the input array \ctext/{I[N]} and
the output array \ctext/{acpt[N]} as \ctext/{N} concurrent ports
instead of the default memory-array interfaces.

In the function body, the per-input local
variables \ctext/{F}, \ctext/{okX} and \ctext/{frontX} also become
arrays parameterized by \ctext/{N}.  We use a fixed-bound
\ctext/{N}-iteration loop to scalably describe the operations
that were previously repetitiously specified for each non-indexed
input \cI1/ and \I2/.  Please note, without the \ctext/{HLS UNROLL} pragma (lines 12 and 19), Vivado-HLS will execute the loop sequentially
over multiple clock cycles.  

%This is how we support structural parameterization when modeling
%routers with, for example, a parameterized number of input ports and
%output ports.

\mypar/{Discussions} 
By elaborating on the same constructions in this section, we have been
able to fully replicate the router designs generated by CONNECT, with
its full range of parameterizations and features.  With careful control
of timing and structure, the routers produced through Vivado-HLS are
exact replacements of their CONNECT counterparts, bit-for-bit,
cycle-for-cycle at the module interface.  Moreover,
Section~\ref{sec:results} will present evaluation results that show
the RTL and HLS counterparts are comparable in their resource
utilization and critical path delay after place-and-route.

There is no magic in Vivado-HLS.  We have been able to arrive at the
desired structural design only by introducing the structural
information explicitly.  Syntax aside, the C functions for the switch
examples above (and also for the CONNECT routers) are strongly
``register-transfer'' in nature.  However, in the design study, we
have found developing and debugging a structural design is
simpler and faster in the Vivado-HLS design flow than
using a conventional RTL flow.  We believe a major reason is the ability
to better separate the concerns for functionality and for performance
(as effected through timing and structure) in the Vivado-HLS design
flow.

Paying no attention to structural consequences, the example C
functions we saw in this section (and same for the CONNECT routers)
have natural sequential program readings of the intended functional
behaviors.  As such, we were able to rely on the convenience of C
testbenches and C debugging tools to first establish a high degree of
confidence in a design's functional correctness before separately
addressing structural and performance design issues.  The
specifications of functionality and structure in conventional RTL
descriptions are too deeply intertwined to do this effectively.

As a final point, we believe our C code for the CONNECT routers would
not be out-of-place in a software cycle-based NoC simulator.  On the
other hand, we do not expect a C router model written expressly for a
software simulator---without consideration for synthesis
implications---would synthesize to the modeled structure or even be
synthesizable at all.  While there are many ways to express the same
behavior in C, not all of them lead to efficient implementations; it
is neither the case in normal C compilation nor in HLS.

\section{Composing Modules in C}
\label{sec:noc}

The Verilog modules of the routers we generated using Vivado-HLS
can be readily instantiated in the next enclosing design hierarchy
using standard structural design methodologies, whether textual or
graphical.  In fact, we could rely on CONNECT to generate the desired
NoC topology as a Verilog netlist and substitute the router modules
from this work for the CONNECT generated counterparts.  Bolstered by
the positive experience in implementing routers using C and Vivado-HLS, our design study next tried to implement the 4-by-4 2D-Mesh in
Figure~\ref{fig::2dmesh}.

\begin{figure}
\centering
\includegraphics[width=0.3\textwidth]{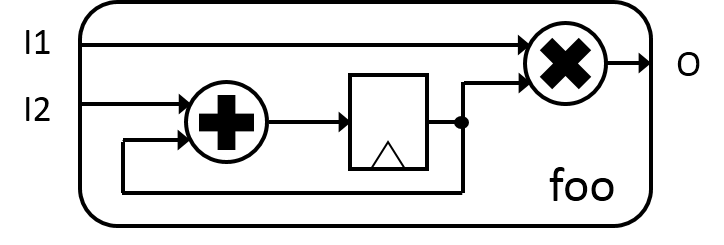}
\caption{A prototypical Mealy state machine.}
\label{fig::foo}
\vspace{-8pt}
\end{figure}

This seemingly benign task---making a netlist of multiple submodules---turned
out to face much greater resistance from C's semantics.  In this
section, we discuss the issues and the workarounds in a general
context with the help of a prototypical Mealy state machine module
(Figure~\ref{fig::foo}).  This module \ctext/{foo} has two integer
inputs \cI1/, \I2/ and an integer output \Oo/. Inside the module,
there is an integer register that accumulates the sequence of values
presented on \I2/ each cycle.  The output \Oo/ is the accumulated sum
scaled by \cI1/.  The corresponding C function \ctext/{foo()} is below.

\begin{lstlisting}
void foo(int I1, int I2, int *O) {
  static int L=INIT_VAL;  // latch
  
  *O=I1*L; // read current-L
  L=I2+L;  // assign next-L 
}
\end{lstlisting}

\begin{figure}
\centering
\includegraphics[width=0.48\textwidth]{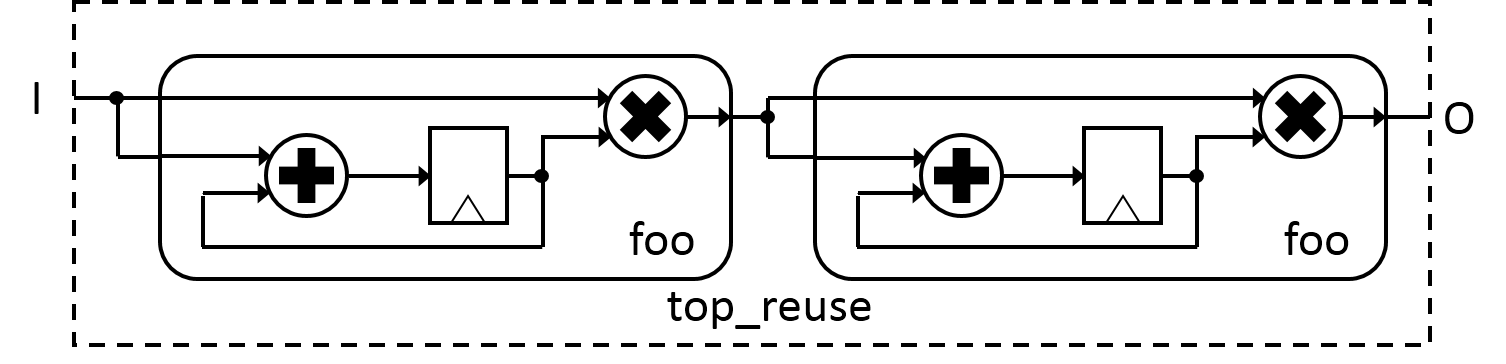}
\caption{Top module \ctext/{top\_reuse} with two instances of the submodule \ctext/{foo}.}
\label{fig::reuse}
\end{figure}

\mypar/{Function Call $\neq$ Module Instantiation} Based on Section~\ref{sec:hls}, it should be clear submitting \ctext/{foo()} as the top-level
function to Vivado-HLS produces the desired module \ctext/{foo} in
Figure~\ref{fig::foo}.  What happens if we want to instantiate two
copies of the module \ctext/{foo} as submodules within the top-level
module \ctext/{top\_reuse} in Figure~\ref{fig::reuse}.  Since
functions and modules both serve as the vehicle for design modularity
and reuse in their respective domains, a naive attempt at describing
the module \ctext/{top\_reuse} might produce the following, where the
function \ctext/{foo()} is called twice.

\begin{lstlisting}
void fxn_reuse_try(int I, int *O) {
  int tmp; // output of left module
  
  foo(I,I,&tmp);  // left in figure
  foo(tmp,tmp,O); // right in figure
}
\end{lstlisting}

Unfortunately, calling a C function twice results in two executions of
the same function instance; it does not result in two copies of the
function.  The use of static variable \Ll/ in \ctext/{foo()} makes
this distinction inescapable.\footnote{Note,
  in Vivado-HLS, multiple calls to a purely combinational function (no
  side-effect through static or global variables) from a combinational
  context will result in replicated instances.  Multiple calls to a
  purely combinational function from a sequential context will result
  in a single instance reused over different clock cycles.  }
  Under C semantics, there is only one
instance of \ctext/{foo()} and one instance of the static variable
\Ll/.  The repeated calls to \ctext/{foo()} {\em{must}} update the
same static variable \Ll/ according to the semantics of C.

\begin{figure}
\centering
\includegraphics[width=0.48\textwidth]{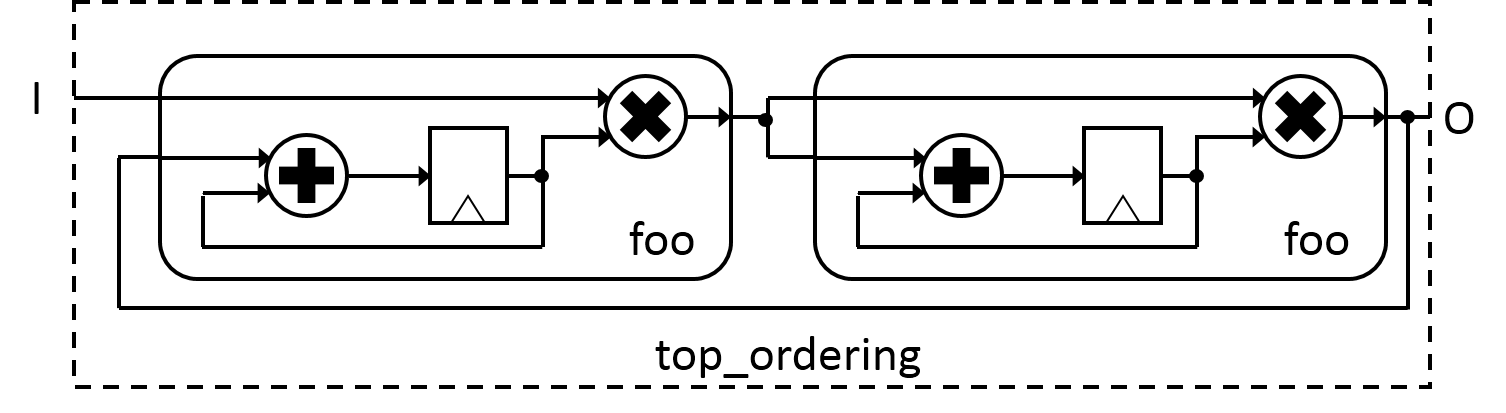}
\caption{Top module \ctext/{top\_ordering} with two instances of the
  submodule \ctext/{foo} where the output of each submodule drives the
  input of the other.}
\label{fig::ordering}
\vspace{-8pt}
\end{figure}

\mypar/{Function Evaluation not Reactive} Consider next the top-level
module \ctext/{top\_ordering} in Figure~\ref{fig::ordering} again with
two instances of the submodule \ctext/{foo}.  This time, the input
\I2/ of the left module is driven by the output \Oo/ of the right
module.  Please note that no combinational cycle is formed. A naive
attempt at a corresponding C function might lead to the following.

\begin{lstlisting}
void fxn_ordering_try(int I, int *O) {
  int tmp1; // output of left module
  int tmp2; // output of right module
  
  foo<1>(I,tmp2,&tmp1);      // left in figure
  foo<2>(tmp1,tmp1,&tmp2);  // right in figure
  *O=tmp2;
}
\end{lstlisting}

We use templatized function with different instance names {\ctext/{foo<1>} and
  {\ctext/{foo<2>}} to workaround the instantiation confusion.
  However, this example highlights a deeper problem arising from that,
  in the module \ctext/{top\_ordering}, each submodule's output \Oo/
  drives the input of the other submodule. There is no ordering of
  calling the functions \ctext/{foo<1>} and \ctext/{foo<2>} in
  \ctext/{fxn\_ordering\_try()} that can produce the desired behavior.
  In the ordering chosen by \ctext/{fxn\_ordering\_try()},
  \ctext/{foo<1>} is called with un-initialized variable \ctext/{tmp2}
  as input; \ctext/{tmp2} is not set until after \ctext/{foo<2>} is
  finished.

It may appear the impasse is caused by a cycle in the netlist. The
real culprit is more generally the order of evaluation.  A
combinational value in a logic circuit should re-evaluate
spontaneously in {\em{reaction}} to changes in its dependent
values. In Verilog/VHDL/SystemC, combinational evaluation is modeled
either as {\em{continuous assign}} statements or as processes whose
sensitivity list contains all the dependent signals.  For reactive
evaluations in Verilog/VHDL/SystemC, the order of declaration does
not matter; the evaluation order develops dynamically, sometimes
including extraneous glitches and redundant evaluations.

The function \ctext/{fxn\_ordering\_try()} looks like---but is not---a netlist. 
The functions \ctext/{foo<1>} and \ctext/{foo<2>} are
evaluated only when they are called and in the order they are called.  The
function \ctext/{foo<1>} is not re-evaluated automatically after \ctext/{tmp2} is
set by \ctext/{foo<2>}.  Explicitly calling \ctext/{foo<1>} one more time is also incorrect because
\ctext/{foo<1>} would make an erroneous extra update to its internal
state \Ll/.  In Verilog/VHDL/SystemC, all synchronous state updates
take place at exactly the same moment as the final act in a clock
period.

\mypar/{Using C++ Objects for Submodules} The alternative to using functions is to
use C++ objects---supported by Vivado-HLS---for modular design and
reuse. The notion of object construction better matches that of
module instantiation.  The module \ctext/{foo} in Figure~\ref{fig::foo}
can be captured as the object class \ctext/{foo\_class} below.

\begin{lstlisting}
class foo_class{
public:
  int L; // latch
  
  foo_class() {L=INIT_VAL;} // class constructor
  void query(int I1, int *O) {*O=I1*L;} // output logic
  void update(int I2) {L=I2+L;}// next-state update
};
\end{lstlisting}

Using this class, we can describe the top-level module in
Figure~\ref{fig::ordering} as the function \ctext/{top\_ordering()}
below.  As in Section~\ref{sec:hls}, the intention is the module
synthesized from this function will execute one invocation completely once per cycle.  The declaration of static objects
\ctext/{foo1} and \ctext/{foo2} in \ctext/{foo\_class} correctly
convey the notion that we want two distinct and persistent instances of the
object \ctext/{foo\_class}.

\begin{lstlisting}
void top_ordering(int I, int *O) {
  static foo_class foo1, foo2;

  int tmp1; // output of left module
  int tmp2; // output of right module

  /* ---- combinational-query behaviors ---- */
  foo1.query(I,&tmp1); 
  foo2.query(tmp1,&tmp2); 
  *O=tmp2; 
  
  /* ---- state-update behaviors ---- */
  foo1.update(tmp2);
  foo2.update(tmp1);
}
\end{lstlisting}

Notice, in \ctext/{foo\_class}, we decomposed the module's behavior
into two methods: a ``query'' method for the output logic and an
``update'' method for next-state update.  This is a part of a
necessary self-imposed coding discipline in our solution. In general,
a submodule object class can have multiple member variables as well as
multiple query and update methods.  A query method output must be purely
combinational (i.e., cannot have any side-effect through member or
global variables and can only be a function of its input
variables and/or member variables).\footnote{Note it is okay to
  call the same query method multiple times; each call results in a
  different combinational logic instance.}  Only an update method can
update the member variables of an object; an update method
cannot have outputs.

\mypar/{Evaluation Order Discipline} Unlike in a netlist where
declaration order does not matter, C++ method invocations must be ordered
deliberately.  For a small example like \ctext/{top\_ordering()},
we can ascertain by inspection that the chosen ordering of method
invocations does lead to the desired behavior. (Vivado-HLS would in
fact produce the correct structure.)  In general, an ordering
discipline is needed for correctness and synthesizability.

The order of query method invocations across all objects must obey
data dependence---before invoking a query method, its input arguments
that are not top-level inputs or static state variables must first be
assigned by an earlier query method.  Since combinational cycles
through query methods are disallowed, a valid dataflow order must
exist. All valid dataflow orderings should result in the same
synthesis outcome.  It is interesting to point out that this same
dataflow ordering requirement is almost second-nature for the 
procedure C code shown in Section~\ref{sec:hls}.

An update method can only be called after all query methods depending
on any of the affected member variables have been called.  The
correct ordering between different update methods of the same object is more
subtle and in general depends on what are the effects of the update
methods and how they are coded.  Alternatively, we propose a general
discipline that restricts an object to a single update method that
updates all member variables in an object.  This update method
must be the last method invoked on the object, that is, after all
query methods have been called.  This general discipline is
stricter than necessary but does not require knowledge of an object's
internal.

\mypar/{Discussions} Using C++ objects and the above ordering
discipline, we can describe and synthesize arbitrary structural
designs comprising synchronous registers and combinational logic using
Vivado-HLS.  This approach is extensible
to a hierarchy of modular objects.  On the downside, one can imagine
the ordering discipline can be cumbersome in larger,
more elaborate designs.  For the design study, while we were able to
describe and synthesize the 2D-Mesh NoC as well as a number of other
common topologies, it would have been much easier to use Verilog for
the netlist.  However, the use of objects and the required ordering
discipline is useful and necessary when mixing objects and inline C
code.  Astute readers may have noticed we already made use of objects
in \ctext/{switch\_buffered()} in Section~\ref{sec:hls} to instantiate
FIFO buffers.

A more interesting question is why is the experience so different when
describing the structure of a router versus the structure of a
NoC. The crux of the answer lies in the fundamental differences in the
semantics of a C function and its invocation versus a module and its
instantiation. Using a function to capture a purely-combinational
module is natural.  We also showed using a function to capture a
top-level module is effective in Section~\ref{sec:hls}.  Keep in mind,
however, when working with the complexity and regularity of a router,
we naturally created a shallow design hierarchy.  We mainly used
sub-functions to encapsulate combinational logic (e.g.,
\ctext/{allocate()} and \ctext/{decode()}).  In
\ctext/{switch\_buffered()} where we called the stateful sub-function
\ctext/{switch\_2stage()}, it was only called once---so there is no
question of multiple instances and a valid evaluation ordering can
be easily built around it.  In contrast, in this section, in a more
general usage of functions to capture sequential submodules, any
illusion that C functions are like hardware modules is inescapably
broken~\cite{Edwards2006}.

\section{Limitations}
\label{sec:tuning}

Using the approaches in Section~\ref{sec:hls} and~\ref{sec:noc}, one
could use C/C++ and Vivado-HLS to describe and synthesize arbitrary
structural designs of synchronous registers and combinational
logic. However, limitations exist when attempting to replicate RTL
designs that make use of macro storage primitives, like FIFOs and
memory blocks.  First, the primitives available for instantiation and inference in C
vs. RTL could be different in structure and in timing behavior.
Second, the rules for inferring primitives are different in C vs. RTL
synthesis.  When facing these differences, sometimes it is possible to
still attain the same bit-level and cycle-level behavior by emulating
the desired but unavailable primitive with an available one.\footnote{In the worst case, one can always resort to emulating the
  unavailable primitive using registers and combinational logic.}  This
emulation will add an overhead cost in logic resources and logic delay.

Consider for example a memory block with one asynchronous read port
and one synchronous write port.  Many suitably formed Verilog
descriptions of this memory block can be automatically mapped to
LUT-RAM by Xilinx RTL synthesis tools.  Surprisingly, we could not
find a way to capture such a basic memory primitive through C and
Vivado-HLS.  The following function conveys the correct intention if
the synthesized module would execute one function invocation 
completely in each cycle, as in all of the previous examples.

\begin{lstlisting}
void ram(int raddr, int *rdata, int waddr, int wdata){
   static int X[8];

   *rdata=X[raddr];
   X[waddr]=wdata;
}
\end{lstlisting}

There is no ostensible reason why this should not work.\footnote{In fact, the desired dual-port asynchronous RAM can be implemented by specifying pragma \ctext/{\#pragma
  HLS RESOURCE variable=X core=RAM\_2P\_1S} in earlier Vivado-HLS verion, for example version 2013.2. However, version 2015.2 we used in our work does not support the storage core RAM\_2P\_1S any more.}  For version 2015.2, Vivado-HLS
does map \ctext/{X} to LUT-RAM, but even with strong prodding using
various pragmas, Vivado-HLS always treats the array read on line 4 as a
synchronous read---resulting in a module that corresponds to a memory
block with a synchronous read port.  If required, one way to achieve the
desired asynchronous read timing is to add the pragma \ctext/{\#pragma
  HLS ARRAY\_PARTITION variable=X complete dim=1} to force the array
\ctext/{X} to be implemented using flip-flops.

Without access to memory with asynchronous read, we could not
replicate an optimization in CONNECT that packs one input port's
multiple virtual-channel flit buffers (only on the data payload
portion\footnote{Among all the virtual channels of one input port, at
  most one flit is added and removed per cycle.  Thus, the data
  payload portion of one input port's multiple virtual-channel flit
  buffers can be time-multiplexed on to one physical structure. The
  route information portion cannot be similarly packed because the
  allocation logic needs to examine the route information from all
  virtual channels in each cycle.})  onto a single random-access memory
structure. In HLS, we have to map an input port's multiple flit
buffers to separate structures. Thus, our HLS-synthesized routers use
more logic to steer the flow to/from the multiple flit-buffer
structures.  This partially accounts for the higher LUT usage that the
HLS-synthesized routers consume relative to CONNECT routers in the
synthesis results in the next section.

On the other hand, for the route information portion of the
virtual-channel flit buffers, the allocator logic needs to read the
first flit of a non-empty flit buffer before deciding whether to
dequeue the flit (i.e., only when the flit is selected to advance).
In the design study, we implemented the flit buffers using
Vivado-supplied \ctext/{ap\_shift\_reg} class which allows this
behavior (reading-before-dequeuing).  In CONNECT, they used a FIFO
primitive that requires dequeuing first before the dequeued flit can
be read in the next cycle.  Both types of FIFO primitives use LUT
for storage so they have equal storage efficiency.  However, in
CONNECT, they have to add an interface shim with flip-flops to convert
from dequeuing-before-reading to reading-before-dequeuing.  This is
why our HLS-synthesized routers consistently use significantly fewer
flip-flops than the corresponding CONNECT routers in the
synthesis results in the next section.\footnote{The CONNECT
  generator could be re-engineered to use SRL16E shift-register FIFOs
  to nullify this difference.}

Finally, in replicating the CONNECT router RTL designs, we replicated
the combinational logic only in the truth-table sense.  We made no
attempt to match up the combinational logic at any more concrete levels
of specification.  As we will see, this can result in small but
inexplicable differences in LUT usage and critical path delay between
our HLS-synthesized routers and the CONNECT routers.

\section{Evaluation and Analysis}
\label{sec:results}

The previous section discussed why the HLS routers are not exactly the
same as the CONNECT routers in the lowest RTL design details.  These
differences have consequences in final implementation
quality---resource usage and critical path delay---when mapped onto
the FPGA fabric.  In this section, we compare the place-and-route
implementation quality of the HLS generated routers
(Section~\ref{sec:hls}}) and NoCs (Section~\ref{sec:noc}) against those
  produced by CONNECT.  The comparison shows the HLS structural
  design methodology can produce comparable quality as CONNECT Verilog
  RTL designs.

\newpage

\begin{center}
\captionof{table}{HLS routers place-and-route ratios over CONNECT routers.}
\label{table::ratio}
\begin{adjustbox}{center}
\scriptsize
%\captionof{table}{HLS routers place-and-route ratios over CONNECT routers.}
%
%\begin{table*}[ht]

    \begin{tabulary}{6cm}{|K{0.4cm}|K{0.4cm}|K{0.35cm} K{0.35cm} K{0.35cm} K{0.35cm}|K{0.35cm} K{0.35cm} K{0.35cm} K{0.35cm}|K{0.35cm} K{0.35cm} K{0.35cm} K{0.35cm}|K{0.35cm} K{0.35cm} K{0.35cm} K{0.35cm}|}
        \hline
        \multicolumn{2}{|c|}{Data width} & \multicolumn{8}{|c|}{32 bits} & \multicolumn{8}{|c|}{128 bits}\\
        \hline
        \multicolumn{2}{|c|}{Num. VCs} & \multicolumn{4}{|c|}{2 VCs} & \multicolumn{4}{|c|}{4 VCs}  & \multicolumn{4}{|c|}{2 VCs} & \multicolumn{4}{|c|}{4 VCs} \\
        \hline
        \multicolumn{2}{|c|}{Buf. Depth} & 4 & 8 & 16 & 32 & 4 & 8 & 16 & 32 & 4 & 8 & 16 & 32 & 4 & 8 & 16 & 32 \\
        %\hline
        %1 & 2 & 3 & 4 & 5 & 6 & 7 & 8 & 1 & 2 & 3 & 4 & 5 & 6 & 7 & 8 & 9 & 9\\
        \hline
        2 & LUT & 0.95 & 0.97 & 0.87 & 0.93 & 1.24 & 1.17 & 0.96 & 0.91 & 1.53 & 1.40 & 1.28 & 1.44 & 2.05 & 1.93 & 1.43 & 1.23\\
        \cline{2-2}
        I/O & FF & 0.44 & 0.43 & 0.38 & 0.37 & 0.41 & 0.48 & 0.38 & 0.35 & 0.44 & 0.43 & 0.38 & 0.37 & 0.41 & 0.42 & 0.35 & 0.35\\
        \cline{2-2}
        Ports & CP & 0.85 & 1.01 & 0.97 & 1.03 & 0.97 & 0.90 & 0.87 & 0.98 & 0.92 & 1.09 & 1.05 & 1.06 & 1.04 & 1.00 & 0.98 & 0.95\\
        \hline
        4 & LUT & 1.08 & 1.05 & 1.12 & 1.01 & 1.25 & 1.19 & 1.10 & 1.05 & 1.72 & 1.60 & 1.76 & 1.64 & 1.80 & 1.69 & 1.46 & 1.29\\
        \cline{2-2}
        I/O & FF & 0.48 & 0.46 & 0.40 & 0.39 & 0.42 & 0.42 & 0.36 & 0.36 & 0.49 & 0.46 & 0.40 & 0.39 & 0.42 & 0.43 & 0.36 & 0.35\\
        \cline{2-2}
        Ports & CP & 0.98 & 1.12 & 0.95 & 1.02 & 0.99 & 0.86 & 0.76 & 0.89 & 0.87 & 0.95 & 0.87 & 0.97 & 0.88 & 0.88 & 0.86 & 0.91\\
        \hline
        6 & LUT & 1.07 & 1.11 & 0.99 & 1.31 & 1.06 & 1.09 & 1.00 & 0.91 & 1.53 & 1.55 & 1.38 & 1.48 & 1.54 & 1.52 & 1.37 & 1.13\\
        \cline{2-2}
        I/O & FF & 0.50 & 0.48 & 0.42 & 0.41 & 0.43 & 0.42 & 0.38 & 0.38 & 0.50 & 0.48 & 0.42 & 0.41 & 0.43 & 0.43 & 0.39 & 0.37\\
        \cline{2-2}
        Ports & CP & 0.77 & 0.96 & 0.84 & 0.87 & 0.81 & 0.82 & 0.82 & 0.82 & 0.79 & 0.80 & 0.79 & 0.83 & 0.88 & 0.77 & 0.75 & 0.84\\
        \hline
        8 & LUT & 1.20 & 1.18 & 1.20 & 1.19 & 1.18 & 1.24 & 1.05 & 1.01 & 1.85 & 1.73 & 1.75 & 1.71 & 1.90 & 1.74 & 1.53 & 1.39\\
        \cline{2-2}
        I/O & FF & 0.54 & 0.52 & 0.45 & 0.44 & 0.45 & 0.44 & 0.39 & 0.38 & 0.54 & 0.52 & 0.45 & 0.46 & 0.45 & 0.44 & 0.39 & 0.38\\
        \cline{2-2}
        Ports & CP & 0.79 & 0.79 & 0.69 & 0.73 & 0.80 & 0.77 & 0.76 & 0.76 & 0.78 & 0.80 & 0.79 & 0.73 & 0.79 & 0.79 & 0.82 & 0.75\\
        \hline
        
    \end{tabulary}
%\end{table*}
\end{adjustbox}
\end{center}

\subsection{Methodology}
The comparison of standalone routers samples a space of configurations
supported by CONNECT.  The space is the cross-product of (a)
in/out-degree (2, 4, 6, 8); (b) flit data width (32-bit and 128-bit);
(c) number of virtual channels (VCs) (2 and 4); and (b) the depth of
flit buffers (4, 8, 16, 32).  All of the routers are configured to use
credit-based flow control, separable input-first allocator and
round-robin arbiter.  All of our HLS router examples are based on a
common parameterized C++ object class configured through
\ctext/{\#define} constants.

The comparison of NoCs samples a range of 16-end-point topologies.
The topologies include Ring, DoubleRing, FatTree, Mesh, Torus,
HighRadix. As with CONNECT NoCs, these different-topology NoCs are
transparently interchangeable in 16-endpoint applications.  The Ring
topology represents a simple low-cost topology; the HighRadix topology
represents a high-cost, high-performance design
point.\footnote{HighRadix is a customized topology with 8 routers
  fully-connected and each router supports two nodes, which means each
  router has 9 I/O ports~\cite{connect_fpga}.} The HLS NoC designs are
hand-coded top-level functions that make use of C++ router objects
following the ordering discipline prescribed in Section~\ref{sec:noc}.
The routers used in this comparison have 2 virtual channels per input
port, 8 flits per flit buffer, 32-bit flit data width.  All of
the routers are configured to use credit-based flow control, separable
input-first allocator and round-robin arbiter.  The above choices
reflect the most commonly chosen router configuration on the CONNECT
website.  The in/out-degree of the routers is topology dependent.

We evaluate implementation qualities in terms of (1) LUT: number of
LUTs consumed; (2) FF: number of flip-flops consumed; and (3) CP: the
clock period.  Neither the HLS routers nor the CONNECT routers use
DSPs or Block-RAMs; flit buffers are implemented using LUTs.  The
reported values are from place-and-route reports targeting a Xilinx
Virtex-7 VX690T FPGA (xc7vx690t, speed grade -2).  The HLS routers
are synthesized using the HLS flow in Xilinx Vivado 2015.2 with the
``evaluate" option (suggested by~\cite{xapp1209} to run both synthesis and
implementation in a single flow).  The Verilog RTL from the CONNECT
generator is synthesized and implemented using the normal RTL Verilog flow in Xilinx
Vivado 2015.2 with the same strategies used in HLS counterparts.  For both HLS and RTL synthesis, for each
configuration, we swept the target clock period (in increments of
1~ns) in repeated runs and selected the result from the synthesis
that yielded the shortest clock period.
 
\subsection{HLS vs. CONNECT}

Table~\ref{table::ratio} gives the ratio of LUT, FF and CP for HLS routers over CONNECT routers.  The HLS routers on average
use 1.33x LUTs, 0.42x FFs, and can achieve 0.87x CP relative
to the corresponding CONNECT routers.  The significant reduction in
flip-flop usage is explained in Section~\ref{sec:tuning}.  In terms of LUT
usage, the HLS routers compare less favorably in the 128-bit wide
configurations than in the 32-bit wide configurations (which are close
to parity with CONNECT routers). This can be attributed to the
differences in the implementations of the virtual-channel flit buffers for holding the data payloads (also explained in Section~\ref{sec:tuning}).  More precisely, the difference is not in the LUTs used
for storage but in the logic LUTs in the datapath surrounding
the different storage structures in use.  In configurations that do not use
virtual channels (not shown in the table), HLS routers are equally
efficient relative to CONNECT routers at 32-bit and 128-bit data
width.

We do not fully understand why the HLS routers can have better critical
path delay than CONNECT routers.  The critical path delay in all cases
is in the allocator combinational logic.  The allocator combinational
logic in the HLS routers and CONNECT routers are identical in the
truth-table sense.  They are specified differently of course, and they
are subjected to different combinational logic optimizations in their
respective synthesis flows.  What is most curious is that the gap
widens with the complexity of the allocator (higher in/out-degree and
more virtual channels). We know this trend is found only in those
configurations using virtual channels.

%%There are too many variables for any conclusions to be drawn without
%%further diagnoses.  For one, we naturally expended great effort in
%%minimizing the critical path delay\footnote{This is actually the very
%%  part of the router that we begin this work on to see how precisely
%%  we can control the synthesis outcome. The final version of our C
%%  description of the allocator logic is effectively a gate-level
%%  netlist}.

Table~\ref{table::network} reports the HLS NoCs place-and-route ratios in comparison to the equivalent CONNECT NoCs.  The
ratios are consistent with the
comparisons of the underlying router modules discussed earlier.  This
affirms that the use of C++ objects and the requirement of the ordering discipline from
Section~\ref{sec:noc} do not have inadvertent consequences.  At this
NoC-level comparison, for the chosen 32-bit-wide router
configurations, the HLS NoCs and CONNECT NoCs are closely comparable
in overall quality.

\begin{table}
    \centering
    \caption{HLS NoCs place-and-route ratios over CONNECT NoCs.}
    \label{table::network}
    \begin{tabular}[b]{|c|c|c|c|} \hline
    Network & $r_{LUT}$ & $r_{FF}$ & $r_{CP}$\\ \hline
    Ring & 1.01 & 0.43 & 1.24\\ \hline
    DoubleRing & 1.20 & 0.44 & 1.07\\ \hline
    FatTree & 1.13 & 0.46 & 1.10\\ \hline
    Mesh & 1.03 & 0.44 & 1.02\\ \hline
    Torus & 1.09 & 0.46 & 1.22\\ \hline
    HighRadix & 1.23 & 0.29 & 0.86\\ \hline
    \end{tabular}
\end{table}

\section{Conclusions}
\label{sec:conclusion}

The answer to the question---\textbf{\textit{can}} we use C and
Vivado-style HLS for structural design---is yes.  We showed that
arbitrary structural designs of synchronous registers and
combinational logic can be captured using the Vivado-HLS methodology.
While there may be limitations when trying to replicate specific macro
storage primitives, most of the time, there are sufficiently
close substitutes to achieve the desired bit-level and cycle-level
behavior with only modest overhead.  We further demonstrated that the
Vivado-HLS methodology is able to produce comparable quality
implementation outcomes as a standard RTL methodology.

The answer to the question---\textbf{\textit{should}} we use C and
Vivado-style HLS for structural design---is sometimes.  The reference
CONNECT RTL designs are generated so we cannot quantify the time and effort
involved.  Quantifying our own design productivity meaningfully is
also very challenging.  To provide one quantitative data point, the
entire effort reported---including the initial learning curve and the
final evaluation analysis---is done by 1 PhD student in 7 months.
More to the point though, as mentioned in the discussions in
Section~\ref{sec:hls}, the perceived productivity gain in the router
portion of the design study is very much noticeable.  We attributed
this gain to the ability to decouple design concerns for functionality
and structure under the Vivado-HLS flow.  However, as later mentioned in the discussions in
Section~\ref{sec:noc}, the Vivado-HLS methodology has a harder time in
handling hierarchical modular structural deigns.  In particular, the
methodology offers little advantage for pure netlisting.  The
break-even for adopting Vivado-HLS in these latter uses is highly
design and context dependent.

%%our NoC design study strongly suggests that the answer is
%%yes. Although currently not all of the hardware details can be
%%replicated, the functionality of our router and NoC are exactly same
%%as their reference---bit- and cycle-accurate at module
%%interfaces. During our exploration, we found that C++ object maps to
%%module better than function in structural design and the order of
%%objects' methods should follow certain discipline. In fact, we believe
%%by using objects and order discipline, we can implement arbitrary
%%synchronous structural design. Though, to get the desired structural
%%design we have to express the structure details in C/C++ code and
%%instruct the tool with pragmas, we still believe it is worth to
%%implement the design at high level. The HLS-generated hardware quality
%%is very comparable with RTL-based design, and designers could benefit
%%from the higher productivity--- better separation of functionality and
%%performance compared with conventional RTL design flow.
%%%\end{document}  % This is where a 'short' article might terminate

\bibliographystyle{abbrv}
\bibliography{sigproc}

\end{document}